\documentclass[aip,rsi,reprint,amsfonts,amssymb,amsmath,floatfix]{revtex4-1}

\usepackage{graphicx}
\usepackage{dcolumn}
\begin{document}

\title{Instrumentation for laser physics and spectroscopy using 32-bit microcontrollers with an Android tablet interface}

\author{E. E. Eyler}
\affiliation{Physics Department, University of Connecticut, Storrs, CT 06269, USA}

\begin{abstract}
Several high-performance lab instruments suitable for manual assembly have been developed using low-pin-count 32-bit microcontrollers that communicate with an Android tablet via a USB interface.  A single Android tablet app accommodates multiple interface needs by uploading parameter lists and graphical data from the microcontrollers, which are themselves programmed with easily-modified C code.  The hardware design of the instruments emphasizes low chip counts and is highly modular, relying on small ``daughter boards" for special functions such as USB power management, waveform generation, and phase-sensitive signal detection.  In one example, a daughter board provides a complete waveform generator and direct digital synthesizer that fits on a 1.5"$\times$0.8" circuit card.

\end{abstract}
\maketitle

\section{\label{sec:intro}Introduction}
In 2011, I described a timing sequencer and related laser lab instrumentation based on 16-bit microcontrollers and a homemade custom keypad/display unit.\cite{Eyler11}  Since then, two new developments have enabled a far more powerful approach: the availability of high-performance 32-bit microcontrollers in low-pin-count packages suitable for hand assembly, and the near-ubiquitous availability of tablets with high-resolution touch-screen interfaces and open development platforms.

  This article describes several new instrument designs tailored for research in atomic physics and laser spectroscopy. Each utilizes a 32-bit microcontroller in conjunction with a USB interface to an Android tablet, which serves as an interactive user interface and graphical display.  These instruments are suitable for construction by students with some experience in soldering small chips, and are programmed using standard C code that can easily be modified.  This offers both flexibility and educational opportunities.  The instruments can meet many of the needs of a typical optical research lab: event sequencing, ramp and waveform generation, precise temperature control, high-voltage PZT control for micron-scale optical alignment, diode laser current control, rf frequency synthesis for modulator drivers, and dedicated phase-sensitive lock-in detection for frequency locking of lasers and optical cavities.  The 32-bit processors have sufficient memory and processing power to allow interrupt-driven instrument operation concurrent with usage of a real-time graphical user interface.

\begin{figure}
\includegraphics[width=\linewidth]{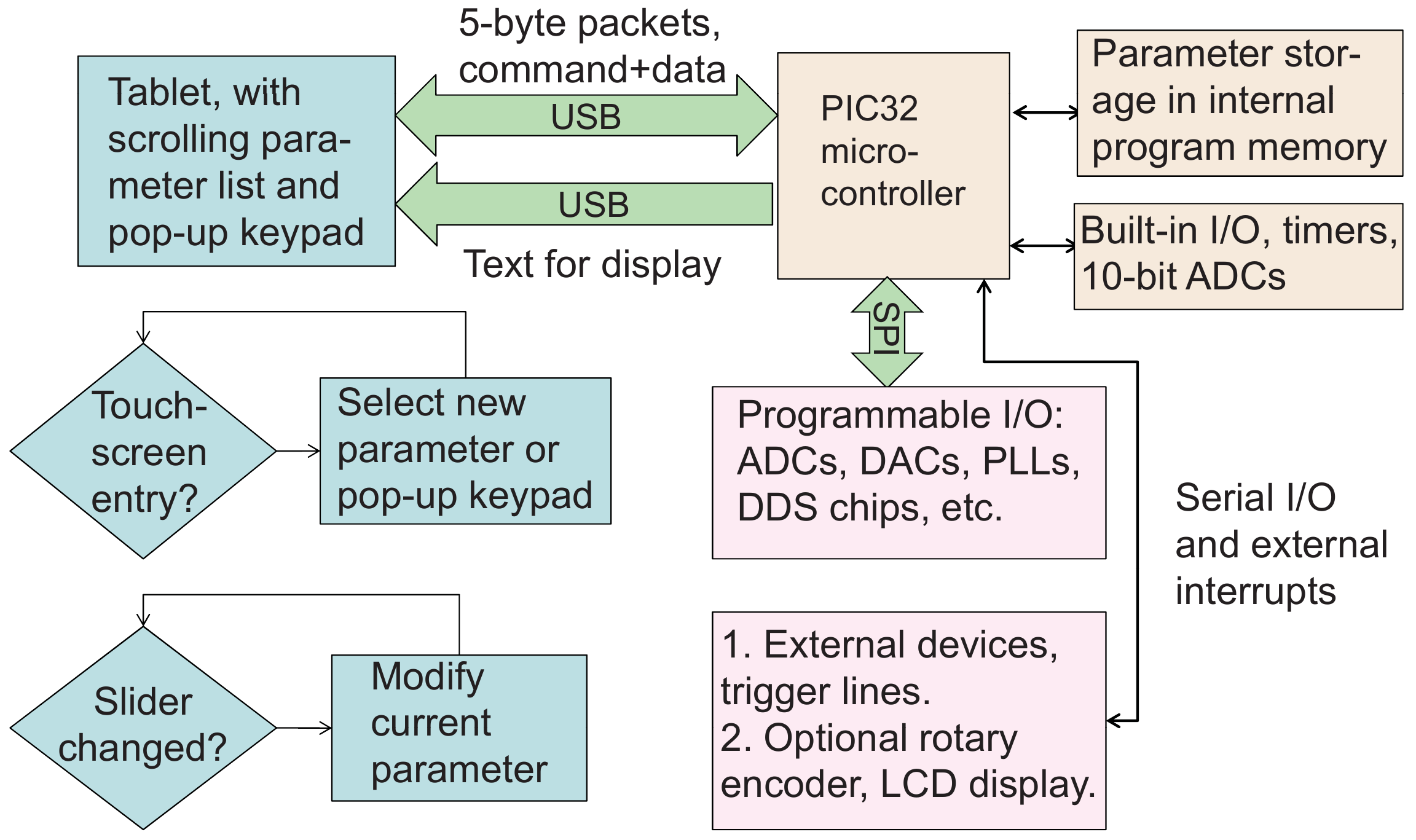}
\caption{\protect\label{SystemOverview} (Color online) block diagram of a microcontroller-based instrument communicating with an Android tablet via USB.  A tablet app, MicroController, uploads parameter values and their ranges from the instrument each time the USB interface cable is connected.}
\end{figure}

 The central principle in designing these instruments has been to keep them as simple and self-contained as possible, but without sacrificing performance.  With simplicity comes small size, allowing control instrumentation to be co-located with optical devices --- for example, an arbitrary waveform synthesizer could be housed directly in a diode laser head, or a lock-in amplifier could fit in a small box together with a detector.  As indicated in Fig. \ref{SystemOverview}, each instrument is based on a commodity-type 32-bit microcontroller in the Microchip PIC32 series, and can be controlled by an Android app designed for a 7" or 8" tablet.  An unusual feature is that the tablet interface is fully interchangeable, using a single app to communicate with any of a diverse family of instruments as described in Sec. \ref{subsec:USB}. Further, all of the instruments are fully functional even when the external interface is removed.  When the operating parameters are modified, the values are stored in the microcontroller program memory, so that these new values will be used even after power has been disconnected and reconnected.  The USB interface also allows connection to an external PC to provide centralized control.

Four printed-circuit boards (PCBs) have so far been designed.  One, the LabInt32 board described in Section \ref{sec:LabInt}, is a general-purpose laboratory interface specifically designed for versatility.  The others are optimized for special purposes, as described in Section \ref{sec:SpecialPurpose}.

The PCBs use a modular layout based in part on the ``daughter boards" described in Sec. \ref{subsec:DaughterBoards}.  They range from simple interface circuits with just a handful of components to the relatively sophisticated Wvfm32 board, which uses the new Analog Devices AD9102 or AD9106 waveform generation chips to support a flexible voltage-output arbitrary waveform generator and direct digital synthesizer (DDS).  It measures 1.5"$\times$0.8", much smaller than any comparable device known to the author.

Further details on these designs, including circuit board layout files and full source code for the software, are available on my web page at the University of Connecticut.\cite{E3web}

\section{\label{sec:system}System Considerations}
In designing the new instrumentation I considered several design approaches.  One obvious method is to use a central data bus, facilitating inter-process communication and central control.  Apart from commercial systems using LabVIEW and similar products, some excellent homemade systems of this type have been developed, including an open-source project supported by groups at Innsbruck and Texas.\cite{Steck09,strontiumlab}  This approach is best suited to labs that maintain a stable long-term experimental configurations of considerable complexity, such as the apparatus for Bose-Einstein condensation that motivated the Innsbruck/Texas designs.

As already mentioned, the approach used here is quite different, intended primarily for smaller-scale experiments or setups that evolve rapidly, where a flexible configuration is more important than providing full central control from a single console.  The intent is that most lab instruments will operate as autonomous devices, although a few external synchronization and control signals are obviously needed to set the overall sequence of an experiment.  These can come either from a central lab computer or, for simple setups, from one of the boards described here, set up as an event sequencer and analog control generator.  This approach is consistent with our own previous work and with recent designs from other small laser-based labs.\cite{Shiell06}

Once having decided on decentralized designs using microcontrollers, there are still at least three approaches: organized development platforms, compact development boards, or direct incorporation of microcontroller chips into custom designs.  Numerous development platforms are now available, ranging from the hobbyist-oriented Arduino and Raspberry PI to more engineering-based solutions.\cite{PI13} However, these approaches were ruled out because they increase the cost, size, and complexity of an instrument.  For simple hardware-oriented tasks requiring rapid and repeatable responses, a predefined hardware interfacing configuration and the presence of an operating system can be more of a hindrance than a help.

Initially it seemed attractive to use a compact development card to simplify design and construction.  My initial design efforts used the simple and affordable MINI-32 development card from MikroElektronika,\cite{MINI32} which combines an 80 MHz Microchip PIC32MX534F064H processor with basic support circuitry and a USB connector.  This board was used to construct a ramp generator and event sequencer very similar in design to an earlier 16-bit version.\cite{Eyler11}  While successful, this approach entailed numerous inconveniences: the microcontroller program and RAM memories are too small at 64 kB and 16 kB, the oscillator crystal is not a thermally stabilized TXCO type, the USB interface requires extensive modification to allow host-mode operation, and the 80 MHz instruction rate is somewhat compromised by mandatory wait states and interrupt latency.  Finally, certain microcontroller pins that are essential for research lab use, such as the asynchronous timing input T1CK, are assigned for other purposes on the MINI-32, requiring laborious cutting and resoldering of traces.  Tests of the event sequencer yielded reasonably good results: the maximum interrupt event rate of 1.5 MHz is about twice as fast as the 16-bit design operating at 20 MHz, although the typical interrupt latency of 400 ns is not very different.  Nevertheless, it became evident that the effort in using preassembled development boards outweighs the advantages.

\begin{figure}
\includegraphics[width=\linewidth]{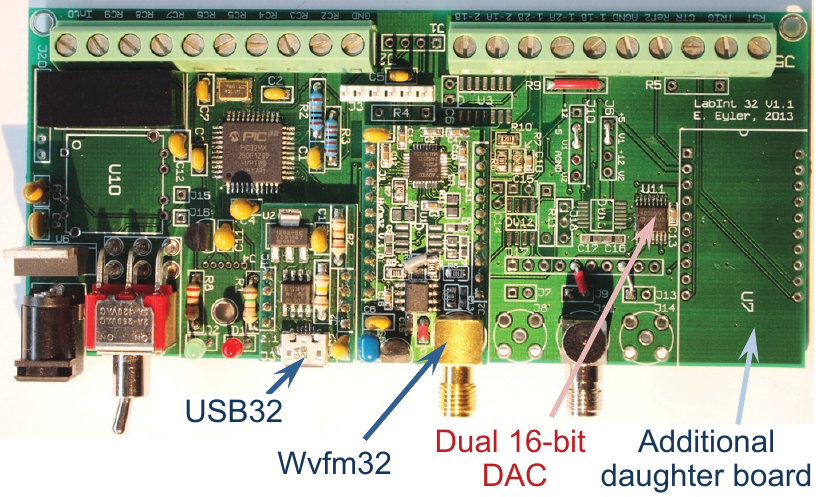}
\caption{\label{LabIntPhoto} (Color online) Photograph of the 5"$\times$2.25" LabInt32 PCB.  It includes a Wvfm32 daughter board with required support circuitry, and a dual 16-bit DAC with one output connected to a card-edge SMA connector.}
\end{figure}
Instead, the designs described here use low-pin-count chips in the Microchip PIC32MX250 series that are directly soldered to the circuit boards, as can be seen in Fig. \ref{LabIntPhoto}.  These microcontrollers, even though they are positioned as basic commodity-type devices by the manufacturer, have twice the memory of the MINI-32 processor and can operate at 40 MHz without wait states.\cite{PIC32Data}  They feature software-reassignable pins that increase interfacing flexibility, as described in Sec. \ref{subsec:modular}.  While the reduced 40 MHz speed is a consideration for event sequencing, it does not impact the performance of any of the other instruments described here, and the absence of wait states during memory access is partially compensatory.  The processor clock and other timing references are derived from miniature temperature-compensated crystal oscillators in the FOX Electronics FOX924B series, which are small, inexpensive, and accurate within 2.5 parts per million.

  Ease of construction is a major consideration for circuits used in an academic research lab.  To facilitate this, the easily-mounted 28-pin PIC32MX250F128B microcontroller is used where possible, and a 44-pin variant when more extensive interfacing is needed.  The basic support circuitry for the controller is laid out to allow hand soldering, as is other low-frequency interface circuitry.  Nevertheless, all of the PCBs include at least a few surface-mounted chips that are more easily mounted using hot-air soldering methods.  We have obtained very good results using solder paste and a light-duty hot-air station.\cite{Aoyue}  For rf circuits the hot-air method is unfortunately a necessity, because modern rf chips commonly use compact flat packages such as the QFN-32, with closely-spaced pins located underneath the chip.  Construction can also be made easier by including a full solder mask on the PCB, greatly reducing the incidence of accidental solder bridges between adjacent pins.  These masks are available for a modest extra fee from most PCB fabricators, and their additional services usually also include printed legends that can conveniently label the component layout.

\subsection{\label{subsec:USB}USB interface to an Android tablet}

\begin{figure}
\includegraphics[width=\linewidth]{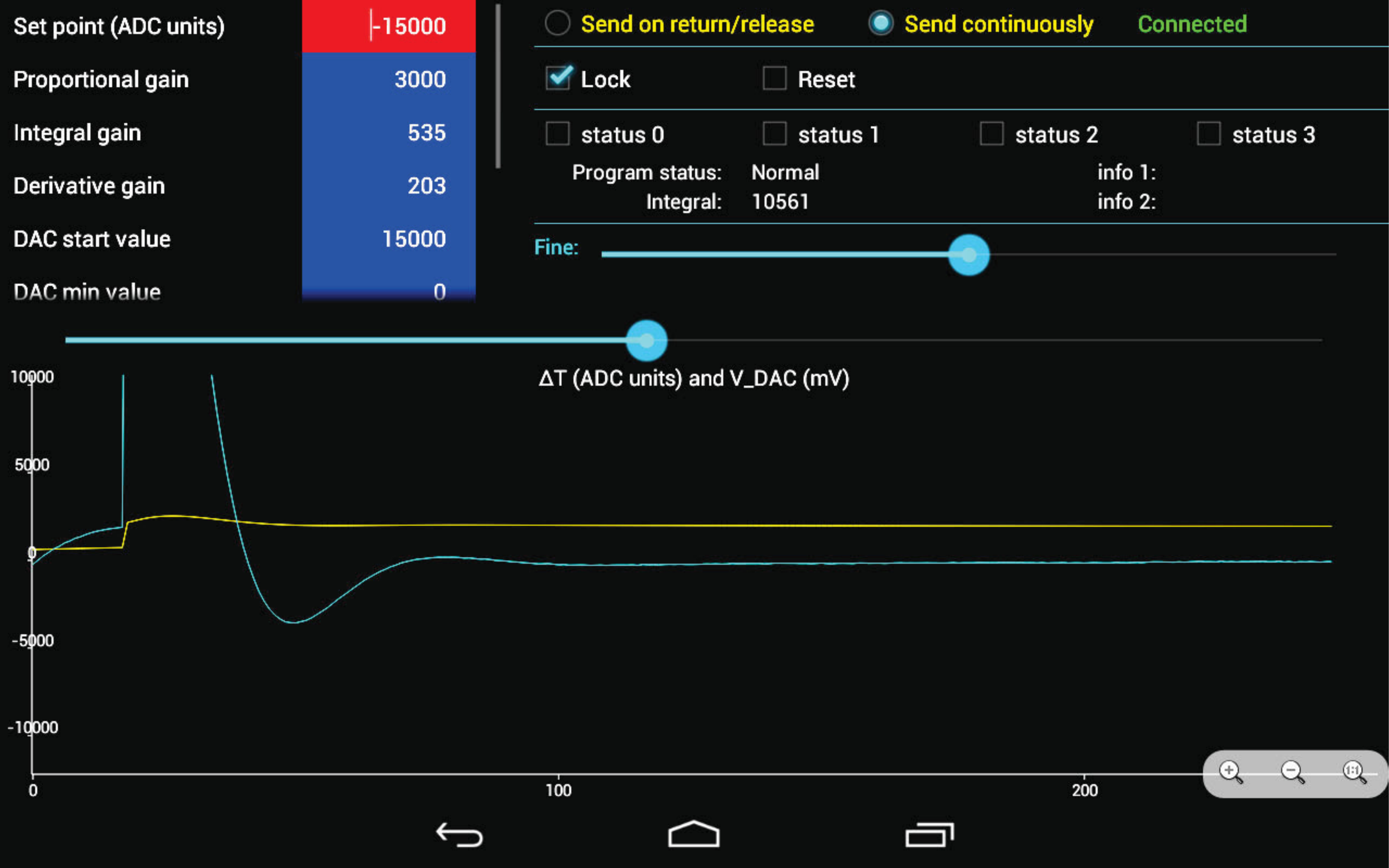}
\caption{\protect\label{ScreenShot} (Color online) Typical screen view of the MicroController app on a Google Nexus 7 tablet.  When a parameter is selected on the scrollable list at the upper left, its value can be adjusted either with a pop-up keypad or with the two slider bars.  The strip-chart graph shows in yellow the output voltage produced by a temperature controller card, and in blue the temperature offset from the set point (25 units $\approx$ 1 mK).}
\end{figure}

As previously described, a user interface to a commodity-type tablet is very appealing because it offers a fast, responsive high-resolution graphical touch-screen interface that requires no specialized instrumentation or construction.  Although rf communication with a tablet is possible using Bluetooth or Wi-Fi protocols, a USB interface is a better choice for lab instruments because it avoids the need for extra circuitry, and it avoids the proliferation of multiple rf-based devices operating in a limited space.  An interface based on an open-source development environment is important, so that programs on both the tablet and the microcontroller can be freely modified for individual research needs.  Fortunately the Android operating system provides such a resource, the Android Open Accessory (AOA) protocol.\cite{AOA}  For this reason, the programs described here were developed for the widely available Google Nexus 7 Android tablet, which offers a 1280$\times$800 display and up to 32 GB of memory, with a fast quad-core processor.  The microcontroller programs use the AOA protocol mainly to transfer five-byte data packets consisting of a command byte plus a 32-bit integer.  They also support longer data packets in the microcontroller-to-tablet direction for displaying text strings and graphics.

 An important consideration is that the USB interface at the microcontroller end of the link must operate in host mode because many tablets, including the Google Nexus 7, support only device-mode operation.  An additional consideration is that for extended operation of the graphical display, a continuous charging current must be provided.  The only way to charge most tablets is via the USB connector, and charging concurrent with communication is only possible if the tablet operates in USB device mode.  On the other hand, it is important that the microcontroller USB interface also be capable of device-mode operation, because when control by an external personal computer is desired, the PC will support only host-mode operation.  For this reason, the full USB on-the-go (OTG) protocol has been implemented in hardware, allowing dynamic host-vs-device switching.  Presently the microcontroller software supports only host-mode operation with a tablet interface, but extension to a PC interface would require only full incorporation of the USB OTG sample code available from Microchip.\cite{USB_OTG}

A more subtle hardware consideration is that both of the tablets I have so far examined, the Google Nexus 7 and Archos 80 G9, use internal switching power supplies that present a rapidly shifting load to the 5 V charging supply.  In initial designs, the 5 V power supply on the microcontroller PCB was unable to accommodate the rapidly switched load, causing fluctuations of $\sim100-200$ mV which then propagated to some of the analog signal lines.  A good solution is to provide a separate regulator for the USB charging supply, operating directly from the same 6V input power that powers the overall circuit card.  With this design there is no measurable effect on the 5 V and 3.3 V power supplies used to power chips on the main circuit board.

A single Android app, MicroController, supports all of the instruments described here by using a flexible user interface based on a scrolling parameter list that is updated each time a new USB connection is established.  It was developed in Java using the Android software development kit, for which extensive documentation is available.\cite{Lee11,Mednieks11,AndroidSDK} The app is available on my web page,\cite{E3web} both as Java source code and in compiled form.  As shown in Figs. \ref{SystemOverview} and \ref{ScreenShot}, the app displays a parameter list with labels and ranges specific to the application.  Several check boxes and status indicators are also available, also with application-specific labels.  Once the user selects a parameter by touching it, its value can be changed using either a pop-up keypad or the coarse and fine sliders visible in Fig. \ref{ScreenShot}.  The remainder of the display screen is reserved for real-time graphics displayed using the open-source AChartEngine package,\cite{AChartEngine}, and can show plots of data values, error voltages from locking circuits, and similar information.  The graphics area can be fully updated at rates up to about 15 Hz.

While certain tasks will eventually require their own specialized Android apps to offer full control, particularly arbitrary waveform generation and diode laser frequency locking, the one-size-fits all solution offered by the MicroController app still works surprisingly well as a starting point.  For a majority of the instruments described here, it is also quite satisfactory as a permanent user interface.

\subsection{\label{subsec:modular}Modular circuit design}

Although this paper mentions seven distinct instruments, they are accommodated using only four PCBs, all of which share numerous design elements as well as a common USB tablet interface.  Multiple instruments can also share a single tablet for user interfacing because it needs to be connected only when user interaction is needed, a major advantage of this design approach.

Another common design element is a 5-pin programming header included on each PCB that allows a full program to be loaded in approximately 10-20 seconds using an inexpensive Microchip PICkit 3 programmer.  The programs are written in C and are compiled and loaded to the programmer using the free version of the Microchip XC32 compiler and the MPLAB X environment.\cite{XC32}  The PIC32MX250 processor family further enhances design flexibility by providing numerous software-reassignable I/O pins, so that a given pin on a card-edge interface terminal might be used as a timer output by one program, a digital input line by another, and a serial communication output by a third.

To avoid repetitive layout work and to further enhance flexibility, several commonly used circuit functions have been implemented on small ``daughter boards" as described in Sec. \ref{subsec:DaughterBoards}.  Two of these daughter boards are visible on the general-purpose lab interface shown in Fig. \ref{LabIntPhoto}, as is an unpopulated additional slot.  Some of these daughter boards simply offer routine general-purpose functionality, such as USB power switching, while others offer powerful signal generation and processing capabilities.

With the exception of the 1"$\times$0.8" USB interface board, the daughter boards measure 1.5"$\times$0.8", and share a common 20-pin DIP connector formed by two rows of square-pin headers.  The power supply and SPI lines are the same for all of the boards, while the other pins are allocated as needed.  These connectors can be used as a convenient prototyping area for customizing interface designs after the circuit boards have been constructed, by wire-wrapping connections to the square pins.

\section{\label{sec:LabInt}General-purpose Laboratory Interface}
\begin{figure*}
\includegraphics[width=\linewidth]{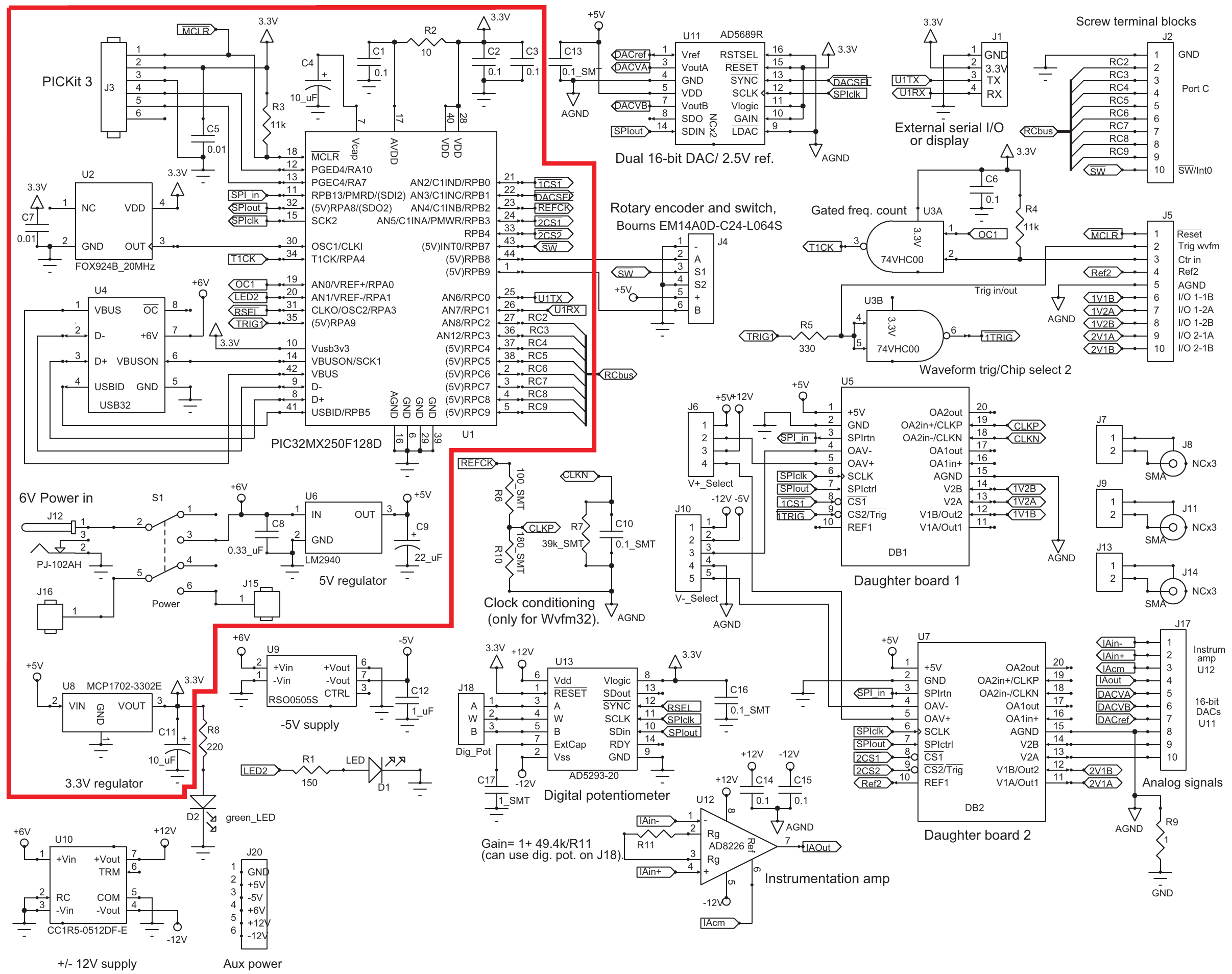}
\caption{\protect\label{LabIntSchematic} (Color online) Schematic diagram for a general-purpose laboratory interface PCB, LabInt32.  Only the portion outlined at the upper left is essential for operation.  The remaining optional components include power supplies, connectors, and a pair of daughter board slots, as well as a dual 16-bit DAC, an instrumentation amplifier, and a 10-bit digital potentiometer.}
\end{figure*}
As already mentioned, the lab interface (LabInt32) PCB was designed to allow a multitude of differing applications by providing hardware support for up to two interchangeable daughter boards, as well as powerful on-board interfacing capabilities.  As shown in Figs. \ref{LabIntPhoto} and \ref{LabIntSchematic}, the core of the design is a PIC32MX250F128D microcontroller in a 44-pin package.  This provides enough interface pins to handle a wide variety of needs, particularly considering that many of them are software-assignable.  Several card-edge connectors and jacks provide access to numerous interface pins and signals, including an 8-bit digital I/O interface, of which six bits are tolerant of 5~V logic levels.  Two of the connectors are designed to support an optional rotary shaft encoder and serial interface as described in Sec. \ref{subsec:CurrentCtrl}.

  The board operates from a single 6 V, 0.5 A power module but contains several on-board supplies and regulators.  These provide the 3.3 V and 5 V power required for basic operation, as well as optional supplies at -5 V and $\pm$12 V for op amps, analog conversion, and rf signal generation.  These optional supplies are small switching power supplies that operate directly from the 6 V input power, so that they do not impose switching transients on the 5 V supply as mentioned in Sec. \ref{subsec:USB}. There are also provisions on the main board for three particularly useful interface components: a dual 16-bit voltage-output DAC with a buffered precision 2.5 V reference (Analog Devices AD5689R), a robust instrumentation amplifier useful for input signal amplification or level shifting (AD8226), and a 1024-position digital potentiometer (AD5293-20) that can provide computer-based adjustment of any signal controllable by a 20 k$\Omega$ resistor, up to a bandwidth limit of about 100 kHz.

Presently there are two demonstration-type programs available for the microcontroller on the LabInt card.  One uses the on-board 16-bit DAC to provide a high-resolution analog ramp with parameters supplied by the tablet interface.  The other operates with the Wvfm32 daughter board to provide a synthesized complex waveform with data output rates up to 96 MHz, as described in the next section.

\subsection{\label{subsec:DaughterBoards}Daughter boards}

The simplest of the daughter boards, the tiny 1"$\times$0.6" USB32 board, is used on all of the PCBs.  It simply provides the power and switching logic for a USB OTG host/device interface, by use of a 0.5 A regulator and a TPS2051B power switch.  It includes a micro USB A/B connector, which is inconveniently small for soldering but is necessary because it is the only connector type that is approved both for host-mode connections to tablets and device-mode connection to external computers.\cite{USB_Spec}  As part of the USB OTG standard, an internal connection in the USB cable is used to distinguish the A (host) end from the B (device) end.

The remainder of the daughter boards are slightly larger at 1.5"$\times$0.8", and they all share a common 20-pin DIP connector as described in Sec. \ref{subsec:modular}. They can be used interchangeably on the LabInt32 PCB or for specific purposes on other PCBs, as for the DAC32 daughter board needed by the TempCtrl card.  This section describes the Wvfm32 daughter board in detail, and briefly describes three others.

\begin{figure*}
\includegraphics[width=0.8\linewidth]{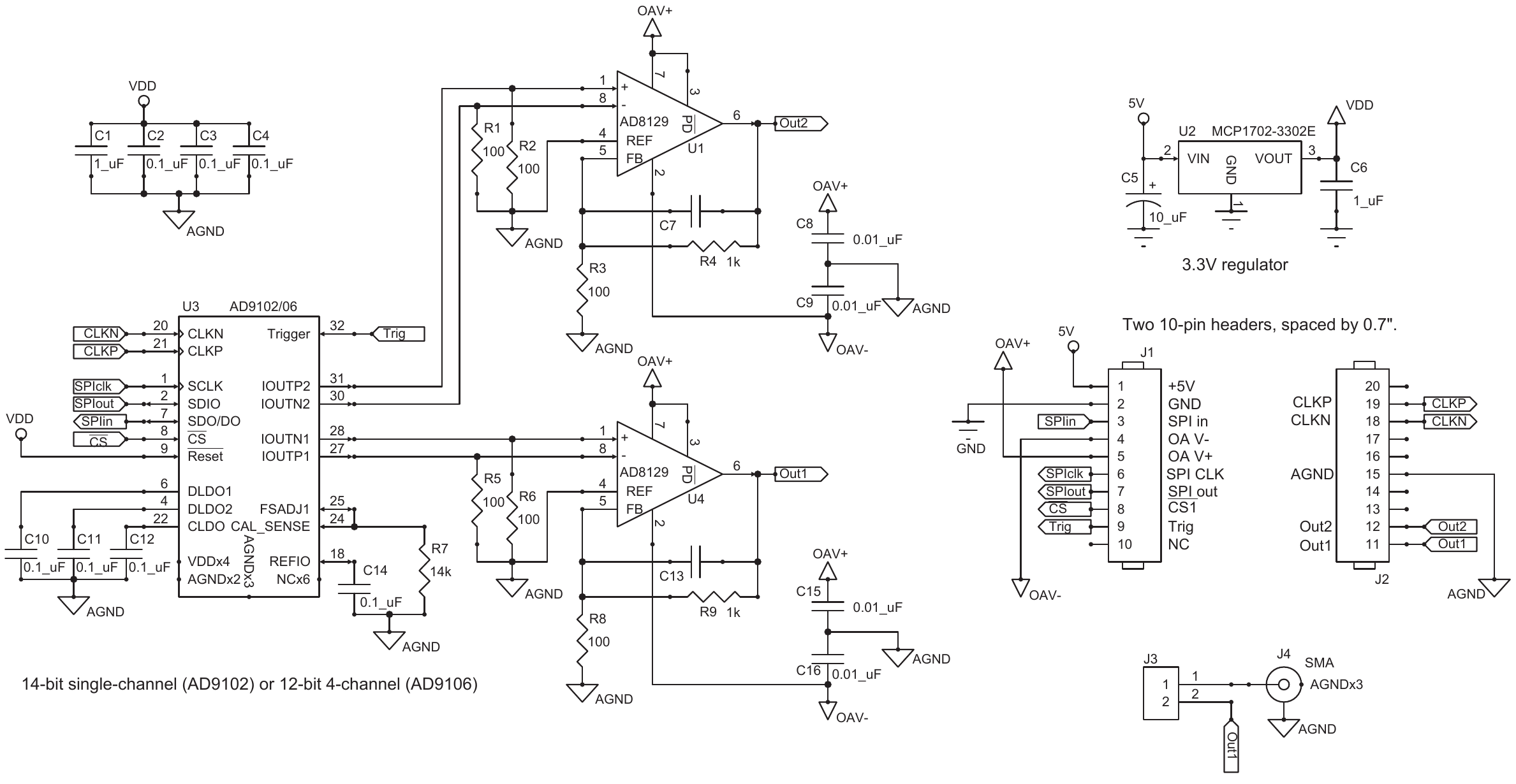}
\caption{\protect\label{Wvfm32} Schematic of a combination arbitrary waveform generator and direct digital synthesizer, using the AD9102 (14 bits, one channel) or AD9106 (12 bits, up to four channels).  The output amplifiers (AD8129 or AD8130) can drive a 50 ohm line with amplitudes exceeding $\pm$2.5~V.}
\end{figure*}
1. \textbf{Wvfm32} The Wvfm32 daughter board, whose schematic is shown in Fig. \ref{Wvfm32}, benefits from the simplicity of a direct SPI interface and provides an extremely small but highly capable instrument.  It combines the remarkable Analog Devices AD9102 (or AD9106) waveform generation chip with a fast dc-coupled differential amplifier (two for the AD9106), along with a voltage regulator and numerous decoupling capacitors necessitated by the bandwidth of about 150-200 MHz.  The AD9102/06 provides both arbitrary waveform generation from a 4096-word internal memory and direct digital synthesis (DDS) of sine waves, with clock speeds that can range from single-step to 160 MHz.  When it is used on the LabInt32 PCB, a fast complementary clock generator is not available, but the programmable REFCLKO output of the PIC32 microcontroller works very well for moderate-frequency output waveforms after it is conditioned by the simple passive network shown near the center of Fig. \ref{LabIntSchematic}.  The REFCLKO output can be clocked at up to 40 MHz using the PIC32 system clock or at up to 96 MHz using the internal USB PLL clock.\cite{PIC32Ref}  Even though the PIC32 output pins are not specified for operation above 40 MHz, the 96 MHz clock seems to work well.

The differential buffer amplifiers, AD8129 or AD8130, can drive a terminated 50 ohm line with an amplitude of $\pm$2.5~V. At full bandwidth the rms output noise level is approximately 1 mV, or 1 part in 5000 of the full-scale output range. The DAC switching transients were initially very large at $\sim$80 mV, but after improving the ground connection between the complementary output sampling resistors (R5 and R6 in Fig. \ref{Wvfm32}), the transients were reduced to 6 mV pulses about 60 ns in duration, and they alternate in sign so that the average pulse area is nearly zero.  The large-signal impulse response was measured using an AD8129 to drive a 1~m, 50~$\Omega$ cable, by setting up the AD9102 waveform generator to produce a step function.  The shape of the response function is nearly independent of the step size for 1--5~V steps.  The output reaches 0.82~V after 4 ns, approaching the limits of the 100 MHz oscilloscope used for the measurement, demonstrating that the circuit approaches its design bandwidth of $\gtrsim 150$~ MHz.  However, it exhibits a slight shoulder after 4 ns, taking nearly 8 ns to reach 90\% of full output and then reaching 100\% at $\sim$11 ns.  After this initial rise, the output exhibits slight ringing at the $\sim$1\% level with a period of about 100~ns, damping out in about three cycles to reach the noise level.  This ringing is caused at least in part by the response of the $\pm$5~V regulators to the sudden change in current on the output line, and is not observed with smaller steps of $\sim$0.1~V.

2. \textbf{DAC32} The DAC32 daughter board is a straightforward design that includes one or two of the same AD5689R DAC chips described in Section \ref{sec:LabInt}, providing up to four 16-bit DAC outputs, together with an uncommitted dual op amp.  The op amps have inputs and outputs accessible on the 20-pin DIP connector, and can be used in combination with the DACs or separately.

3. \textbf{LockIn} The LockIn daughter board does just what its name implies.   It realizes a simple but complete lock-in amplifier, with a robust adjustable-gain instrumentation amplifier (AD8226 or AD8422) driving an AD630 single-chip lock-in amplifier that works well up to about 100 kHz.  The output is amplified and filtered, then digitized by an AD7940 14-bit ADC.  The performance is determined mainly by the AD630 specifications, except that the instrumentation amplifier determines the input noise level and the common-mode rejection ratio.  When used on the LabInt32 PCB, the digital potentiometer on the main board can be tied to this daughter board to allow computer-adjustable gain on the input amplifier.

4. \textbf{ADC32} The ADC32 board is still in the design stage.  It will use an AD7687B 16-bit ADC, together with a robust ADG5409B multiplexer and an Intersil ISL28617FVZ differential amplifier, to provide a flexible high-resolution analog-to-digital converter supporting four fully differential inputs.  In addition to offering higher resolution than the built-in 10-bit ADCs on the PIC32 microcontroller, it offers a much wider input voltage range and considerable protection against over-voltage conditions.
\section{\label{sec:SpecialPurpose}Special-purpose Instrumentation}
\subsection{\label{subsec:TempCtrl}Temperature Controller and PZT driver}

\begin{figure}
\includegraphics[width=\linewidth]{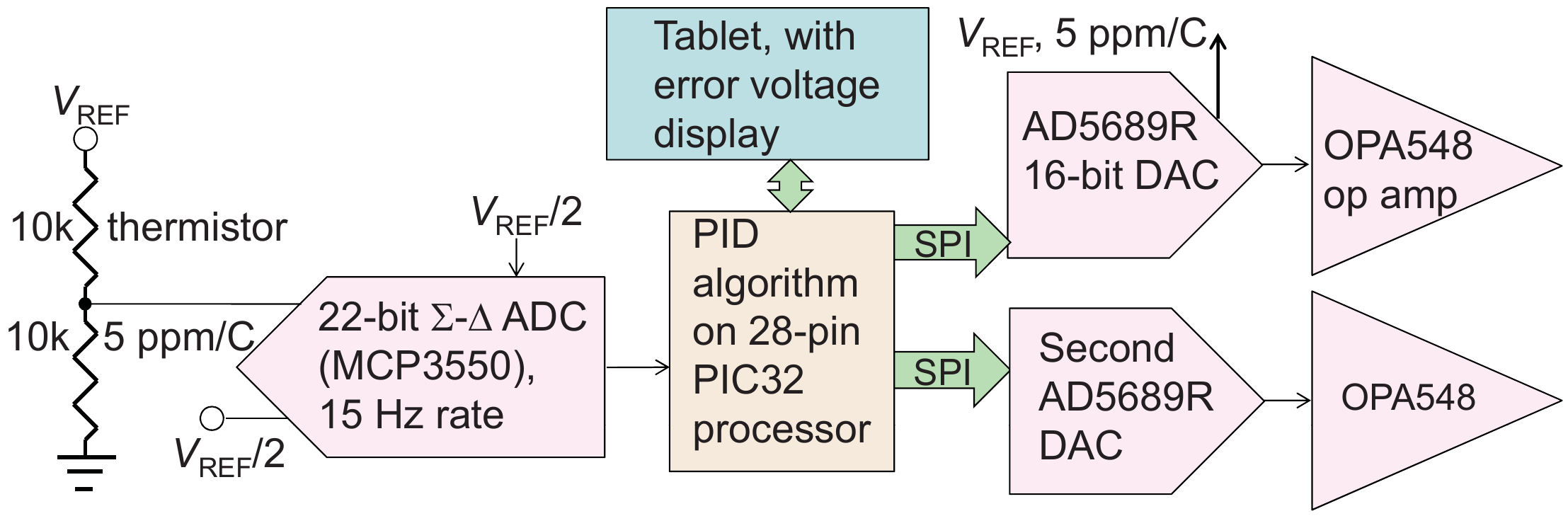}
\caption{\protect\label{TempCtrl} (Color online) Major elements of the precision temperature controller.  High-voltage PA340CC op amps can be substituted for the OPA548 high-current drivers for use as a dual PZT driver.}
\end{figure}

The Temp32 PCB, shown in block form in Fig. \ref{TempCtrl}, uses a 28-pin PIC32MX250F128B on a card optimized specifically for low-bandwidth analog control, with three separate ground planes for digital logic, signal ground, and analog power ground.  While simple enough to be used for general-purpose temperature control, the board was designed to allow the very tight control needed for single-mode distributed Bragg reflector (DBR) lasers, for which a typical temperature tuning coefficient of 25 GHz/C necessitates mK-level control for MHz-level laser stability.\cite{Photodigm}  As shown in Fig. \ref{TempCtrl}, a divider formed by a thermistor and a 5 ppm/C precision resistor provides the input to a 22-bit ADC.  The Microchip MCP3550-60 is a low-cost ``sigma-delta" ADC that provides very high accuracy and excellent rejection of 60 Hz noise at low data rates (\~15 Hz).  A 2.5 V precision reference is used both for the thermistor divider and to set the full-scale conversion range of the ADC, making the results immune to small reference fluctuations. No buffering is required for the thermistor, although if a different sensor were used a low-noise differential amplifier might be desirable.\cite{MCP3550_app_notes}

The microcontroller program implements a PID (Proportional-Integral-Differential) controller using integer arithmetic, with several defining and constraining parameters that can be optimized via the tablet interface.  The output $V$ after iteration $i$ is determined by the error $E_i$, gain factors $G_P$, $G_I$ and $G_D$, the sampling frequency $f$, and a scale factor $S=8192$ that allows the full 16-bit range of the output DAC to be used:
\begin{eqnarray}
&&I_i  =  I_{i-1} + E_i * G_I / f \nonumber \\
&&\text{if } (I_i>I_{\text{max}}) \text{ or } (I_i<I_{\text{min}})\text{, set to limit.} \nonumber\\
&&  \langle D \rangle =  \text{running average of $(E_i - E_{i-1})*f$} \nonumber\\
&&V = V_0 + (E_i*G_P + \langle D \rangle * G_D + I_i )/S \\
&&\text{if } (V>V_{\text{max}}) \text{ or } (V<V_{\text{min}})\text{, set to limit}. \nonumber
\end{eqnarray}
This output is sent a DAC32 daughter board, then amplified by an OPA548 power op amp capable of driving 60~V or 3~A.  The separate analog power ground plane for the output section of the PCB is connected to the analog signal ground plane only at a single point.

With a conventional 10~k$\Omega$ thermistor, the single-measurement rms noise level is approximately 7 ADC units, corresponding to about 0.3 mK near room temperature.  Assuming a bandwidth of about 1 Hz for heating or cooling a laser diode or optical crystal, the time-averaged noise level and accuracy can exceed 0.1 mK, adequate for most purposes in a typical laser-based research lab.

The TEMP32 circuit board can alternatively be used as a dual 350-V PZT controller for laser spectrum analyzers or other micron-scale adjustments.  To accomplish this, the ADC is omitted and the output op amps are substituted with Apex PA340CC high-voltage op amps, using a simple adapter PCB that accommodates the changed pin-out.

\subsection{\label{subsec:FreqSynth}Frequency Synthesizer}
\begin{figure}
\includegraphics[width=\linewidth]{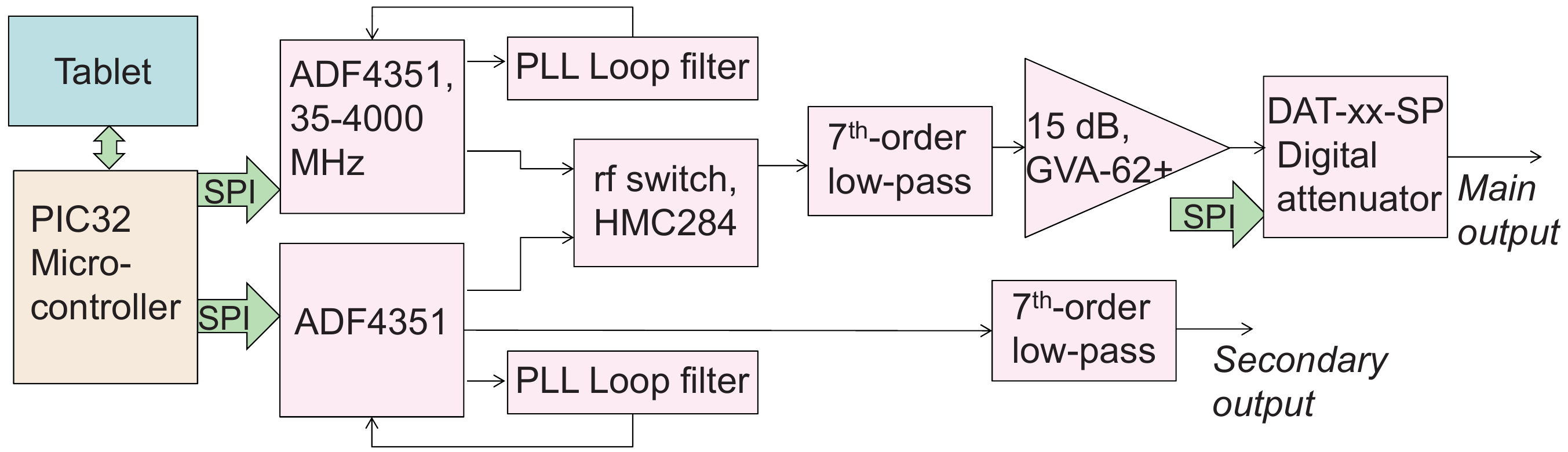}
\caption{(Color online) block diagram of a dual 35-4000 MHz frequency synthesizer, with ns-scale switching and programmable output attenuation.}
\label{FreqSynth}
\end{figure}
The FreqSynth32 PCB, presently in the testing phase, is intended to provide accurate high-frequency rf signals for applications such as driving acousto-optic modulators.  It supports up to two ADF4351 ultra-broadband frequency synthesizers, which can produce far higher frequencies than DDS synthesizers.  These PLL-based devices include internal voltage-controlled oscillators and output dividers, allowing self-contained rf generation from 35-4000 MHz.  As shown in Fig. \ref{FreqSynth}, an rf switch allows ns-timescale switching between the two synthesizers, or if one is turned off, it allows fast on-off switching.  Signal conditioning includes a low-pass filter to eliminate harmonics from the ADF4351 output dividers, as well as a broadband amplifier and digital attenuator that provide an output level adjustable from about -15 dBm to +16 dBm.  The output can drive higher-power amplifier modules such as the RFHIC RFC1G21H4-24, which provides up to 4W in the range 20-1000 MHz.

It is a challenge to work over such a broad frequency range.  Although an impedance-matched stripline design was not attempted because this would require a PCB substrate thinner than the 0.062" norm, considerable attention has been paid to keeping the rf transmission path short, wide, and ``guarded" from radiative loss by numerous vias connecting the front and back ground planes on the PCB.

\subsection{\label{subsec:CurrentCtrl}Laser Current Control Interface}

The MPL\_Interface PCB, described more fully on my web page,\cite{E3web} is designed as a single-purpose interface to a laser diode current driver compatible with the MPL series from Wavelength Electronics.  However, this circuit may be of more general interest for two reasons.  First, it allows control and readout of devices with a ground reference level that can float in a range of $\pm 10$~V, with 13-16 bit accuracy.  Second, its control program includes full support for a rotary shaft encoder (Bourns EM14A0D-C24-L064S) and a simple serial LCD display (Sparkfun LCD-09067), allowing the laser current to be adjusted and displayed without the USB tablet interface.  The same encoder and display could also be attached to the LabInt32 PCB using jacks provided for this purpose.

\section{Conclusions}
A significant portion of the research needs of a typical laser spectroscopy or atomic physics laboratory can be met by the four PCBs described in Secs. \ref{sec:LabInt} and \ref{sec:SpecialPurpose}, together with appropriate software and the daughter boards in Sec. \ref{subsec:DaughterBoards}.  The advantages of this approach include simple and accessible modular designs, a user interface to an Android tablet with interactive high-resolution graphics, and easily reconfigurable software.  The circuit designs are intended for in-house construction, reducing expenses and allowing valuable educational opportunities for students, while still offering the high performance expected of a specialized research instrument.  Most of the PCBs can be hand-soldered, although a hot-air soldering station is required for the two rf circuits (Wvfm32 and FreqSynth32).  Full design information and software listings are available at my website.\cite{E3web}

Apart from these general considerations, these instruments offer some unusual and valuable capabilities.  One is the single shared Android app that provides a full graphical interface to numerous different devices.  When the tablet is removed after adjusting the operating parameters, the microcontroller stores the updated parameter values and the instrument will continue to use them indefinitely.  Another is the very small size of the Wvfm32 waveform generator, which takes advantage of a simple direct interface connection to a microcontroller to provide voltage-output arbitrary waveform generation and DDS on a 1.5"$\times$0.8" PCB.  Up to two of these PCBs can be mounted on a LabInt32 general-purpose interface card, itself measuring only 5"$\times$2.25", and only a single semi-regulated 6V, 0.5A power supply is required.  Similarly, the DAC32 and LockIn daughter boards share the same small footprint, facilitating control instrumentation that can fit inside the device being controlled.

The primary usage of these instruments in our own laboratory is to control several diode lasers and to provide flexible control of numerous frequency modulators needed for research on optical polychromatic forces on atoms and molecules.\cite{Chieda12,Galica13}  Although the available circuits and software reflect this focus, most of these instruments can be used for diverse applications in their present form, and all can be modified readily for special needs.\\

\begin{acknowledgments}
This work was supported by the National Science Foundation.

\end{acknowledgments}



\begin{thebibliography}{21}
\bibitem{Aoyue} Aoyue968A+, available from SRA Soldering Products, Foxboro, MA.  To mount surface-mount chips a thin layer of solder paste with flux should be applied to the PCB pads; we have obtained good results using ChipQuik model SMD291AX.
\end{thebibliography}
\end{document}